\documentstyle[epsfig,12pt]{article}
\title{Neutrino Magnetic Moment Upper Bound From Solar Neutrino Observations}
\author{{\bf Jo\~{a}o Pulido} \\
Centro de F\'{\i}sica das Interac\c{c}\~{o}es Fundamentais \\
Instituto Superior T\'{e}cnico \\
Av. Rovisco Pais, 1096 Lisboa Codex, Portugal\\
{\bf Ana M. Mour\~{a}o} \\
Centro Multidisciplinar de Astrof\'{\i}sica \\
Instituto Superior T\'{e}cnico \\
Av. Rovisco Pais, 1096 Lisboa Codex, Portugal}

\newcommand{\be}{\begin{equation}}
\newcommand{\ee}{\end{equation}}
\newcommand{\bea}{\begin{eqnarray}}
\newcommand{\eea}{\end{eqnarray}}
\newcommand{\gv }{g_{_V}}
\newcommand{\ga }{g_{_A}}
\newcommand{\ph }{P_{_H}}
\begin{document}
\maketitle

\begin{abstract}
    Using the data from SuperKamiokande, Kamiokande and Homestake solar 
neutrino experiments we derive an upper bound on the magnetic moment 
of the neutrino and find $\mu_{\nu_e}\leq(2.2-2.3)\times 10^{-10}\mu_{B}$,
within four different standard solar models. We assume equal magnetic 
moments for all neutrino flavours.
   This limit is obtained when neutrinos do not undergo any "disappearence"
mechanism other than the magnetic moment conversion due to the solar 
magnetic field and for a total or nearly total suppression of the 
intermediate energy  neutrinos.  In our work we consider an energy 
dependent suppression of  solar neutrinos. 
We also point out that the limit may be further reduced if the 
detector threshold energy in $\nu_{e,x}e^{-}$ scattering with 
solar neutrinos is decreased.
\end{abstract}

\newpage

{\bf \large 1. Introduction}

\vspace { 6mm}

The solar neutrino problem, which first appeared as a deficit of the solar
neutrino flux in the Homestake experiment \cite{Davis88} relative to the 
solar model prediction \cite{Bahcall82}, has remained with us since its  
first acknowledgement in the late 1960's. In more recent years the Kamiokande
 \cite{Hirata}, SAGE \cite{Abdurashitov} and Gallex \cite{Anselmann} 
experiments, observing different parts of the neutrino spectrum, started 
operation. Besides these experiments, several theoretical solar models 
\cite{BP95} - \cite{CIR97} have been developed and our understanding of the
situation has changed. It now appears that the solar neutrino problem is not 
merely a deficit of the measured flux in the Kamiokande or the Homestake 
experiment. If it were so, it could be substantially reduced and even 
absorbed within the theoretical uncertainties in the $^{8}B$ neutrino flux
\cite{Crespo}, the only component observed in Kamiokande and the main one
in Homestake. More important, it is the problem of the disappearence of
the intermediate energy neutrinos \cite{Bahcall94} - \cite{B94}. 
This is practically independent of any solar model considerations and 
relies essentially on a detailed analysis of the experimental data on the 
basis of the pp cycle dominance. There are therefore increasingly stronger
indications that the solution to the solar neutrino problem must rely on
non-standard neutrino properties, either neutrino oscillations in matter
\cite{W78}, vacuum \cite{H96},  the magnetic moment \cite{VVO,RSFP}
 or a "hybrid scenario" \cite{big}.

In this paper we aim at establishing a new upper bound on the electron 
neutrino magnetic moment. Our work starts from the analysis of the weak and 
electromagnetic cross section for neutrino electron scattering in the
Kamiokande detector and uses the most recent data from the Homestake 
(Chlorine), Kamiokande and SuperKamiokande experiments.
The first of these experiments is 
looking at a purely weak charged current process, namely
\begin {equation}
\nu_{e}+^{37}{Cl}\rightarrow^{37}{Ar}+e^{-}
\end {equation}
whereas the second is based on elastic scattering,
\begin {equation}
\nu_{e,x}+e^{-}\rightarrow\nu_{e,x}+e^{-}
\end {equation}
with $x=\mu,\tau$  and
where possible electromagnetic properties of the neutrino may play a 
significant role. These are parametrized in terms of the electromagnetic form 
factors which at $q^2\simeq 0$ ammount to the magnetic moment and charge 
radius. We allow for the solar neutrino deficit to be jointly explained in terms
of these electromagnetic effects and any other sources like, for instance,
oscillations. The upper bound on the magnetic moment is of course obtained 
when these other sources are absent. Previous analyses aimed at deriving 
bounds on the neutrino magnetic moment $\mu_{\nu}$ using neutrino electron 
scattering cross sections with electromagnetic interactions exist already in 
the litterature \cite{Suzuki},\cite{MPR}. They did not however include the 
possibility of origins for the solar neutrino deficit other than the magnetic 
moment transition, resulting therefore in upper and lower bounds for 
$\mu_{\nu}$. Furthermore they assumed an energy independent neutrino deficit,
which now appears not to be the case \cite{Bahcall94} - \cite{B94},
\cite{KP}. Several authors \cite{big}
have on the other hand performed investigations using the combined resonant
spin-flip precession mechanism and oscillations which are based on specific 
assumptions on the magnitude and/or solar magnetic field profile and obtain
iso-SNU and survival probability plots for the solar neutrino fluxes. 
The scope of our analysis is however quite 
different, since it takes into account the  neutrino-electron scattering (weak
and electromagnetic) in the Kamiokande detector 
and the actual mechanism suppressing solar electron 
neutrinos $(\nu_{e_L})$ is totally irrelevant here. Further,  we use 
the assumption 
of an energy dependent neutrino deficit as indicated by the combined 
experimental data. Our results are derived for four
different theoretical solar models \cite{BP95,BP92,TCL,TCCCD}. They
show a smooth dependence on $P_{I}$, the survival probability of the 
intermediate energy neutrinos, a parameter which to a very good accuracy
(better than $2\sigma$) can be assumed zero \cite{HR96,C97}. For all 
models we obtain an upper bound in the range $(2.2-2.3)\times {10^{-10}}
{\mu_B}$ for the electron neutrino magnetic moment, an improvement
with respect to the most stringent laboratory bound existing to date,
$\mu_{\nu_e} \leq {10.8\times {10^{-10}} {\mu_B}}$ (90\% CL), from
the LAMPF group \cite{RPP}. More stringent bounds exist, however, for the
electron anti-neutrino magnetic moment at the same order of magnitude of
the numbers obtained here: $\mu_{\bar{\nu}_e} \leq {1.8\times {10^{-10}}
{\mu_B}}$ \cite{Derbin}. Astrophysical and cosmological bounds are on the
other hand even more restrictive. They come from supernova analysis \cite{NR}
$(\mu_{\nu}\leq 10^{-13}\mu_B)$, energy loss in Helium stars \cite{FY} 
$(\mu_{\nu}\leq 8\times {10^{-12}}\mu_B)$, cooling of red giants by plasmon
decay into neutrino pairs \cite{R} $(\mu_{\nu_e}\leq {3\times 10^{-12}}\mu_B)$
and nuclear synthesis in the big bang \cite{M} 
$(\mu_{\nu}\leq 1.5\times{10^{-11}}\mu_B)$. All these bounds cannot be taken 
as literally as the laboratory ones and especially the supernova one \cite{NR}
is considered by some to be avoidable. Morgan's bound \cite{M} can be violated 
by one transition magnetic moment.

We restrict ourselves to the case of Dirac neutrinos. For Majorana neutrinos 
the analysis would be different because an active 
$\bar{\nu}^{\rm M}_{eR}$ could also be present and be detected through the 
process $\bar{\nu}^{\rm M}_ {eR}~ + p\rightarrow n+e^+$ for which there exists
however the firm upper bound from the LSD experiment 
$\Phi_{\bar{\nu}_e}/\Phi_{{\nu}_e} \leq 1.7\%$ \cite{santinu}.
Furthermore the states  $\bar{\nu}^{\rm M}_{\mu,\tau R}$
would now be active under weak interactions. 

The plan of the paper is as follows: in section 2 we describe the method used 
for deriving the upper bound on $\mu_{\nu_e}$ starting from the cross sections 
for $\nu_{e,x} e$ scattering. We discuss its dependence on $P_{H}$, the
survival 
probability for high energy ($^{8}B$) neutrinos which is related to $P_{I}$,
and on $\alpha$ which parametrizes the disappearence due to flavour 
oscillations in each of the available solar models. In this way the present 
paper also differs from our previous work \cite{MPR}, where only essentially 
one model was available and neutrino suppression was considered energy 
independent, affecting equally the $^{8}B$ and the intermediate energy 
($^{7}Be$, CNO) neutrinos which is presently known as not 
being the case. Moreover in \cite{MPR} no matter oscillation effect was 
considered. It is remarkable that, although the $^{8}B$ flux prediction
differs by sizeable amounts for different models with a relative spread of
43\% (see table I), the prediction for the upper bound on $\mu_{\nu}$ using
the recent SuperKamiokande data ranges over a 5\% spread only (see fig. 3).
Finally in section 3 we draw our main conclusions and comment on possible 
future directions.

\vspace{10 mm}

{\bf \large  2. Event Rates and Cross Sections}

\vspace{6 mm}

The event rate in a solar neutrino experiment in which recoil electrons are 
produced is given by the corresponding cross section per unit neutrino energy 
$E_{\nu}$ per unit kinetic energy $T$ of the recoil electron times the neutrino
flux and summed over all possible neutrino fluxes:
\begin{equation}
S_{exp}=\sum_{i} \int dE_{\nu_i} \int \frac{d^{2}\sigma}{dT dE_{\nu_i}} 
f(E_{\nu_i}) dT \label{sexp}
\end{equation}
The quantity $f(E_{\nu_i})$ represents the i-th normalized neutrino flux. 
For Kamio-\\
kande, which is based on neutrino electron scattering, and where only the
$^{8}B$ neutrino flux is seen, we have
\begin{equation}
S_K=\!\int dE_{\nu}\int f(E_{\nu})\left( X_W\frac{d^2\sigma_W}{dT dE_{\nu}}
+\frac{d^2\sigma_{+EM}}{dT dE_{\nu}}+\frac{d^2\sigma_{-EM}}{dT dE_{\nu}}+
X_{int}\frac{d^2\sigma_{int}}{dT dE_{\nu}}\right) dT
\end{equation}
The quantities $X_W$, $X_{int}$ will be derived below. The weak 
($d^2\sigma_W/dTdE_{_\nu}$), electromagnetic spin non-flip 
($d^2\sigma_{+EM}/dTdE_{_\nu}$), 
electromagnetic spin flip ($d^2\sigma_{-EM}/dTdE_{_\nu}$) and interference
($d^2\sigma
_{int}/dTdE_{_\nu}$) parts of the differential cross section were taken 
from \cite{KSN}.
Denoting by $ f_{\nu}$ the neutrino magnetic moment in Bohr magnetons $\mu_B$ we 
have 
\bea
\frac{d^2\sigma_W}{dT dE_{\nu}}&\!=\!&\!\!\!\frac{G^2_{F}m_{e}}{2\pi}
\left((\gv  +\ga )^2+
(\gv -\ga )^2(1-\frac{T}{E_{\nu}})^2-(\gv ^2-\ga ^2)\frac{m_{e}T}{E_{\nu}^2}
\right) \\
~\nonumber\\
\frac{d^2\sigma_{+EM}}{dT dE_{\nu}}&\!=\!&\!<r^2>^2\frac{\pi\alpha^2}{9}m_e
\left( 1+(1-\frac{T}{E_{\nu}})^2-\frac{m_{e}T}{E_{\nu}^2}\right) \\
~\nonumber\\
\frac{d^2\sigma_{-EM}}{dT dE_{\nu}}&\!=\!&\!f_{\nu}^2\frac{\pi\alpha^2}{m_e^2}
\left( \frac{1}{T}-\frac{1}{E_{\nu}}\right) \\
~\nonumber\\
\frac{d^2\sigma_{int}}{dT dE_{\nu}}&\!=\!&\!\!\!-<r^2>\frac{\sqrt{2}}{3}
\alpha G_{F}m_e
\left( \gv \frac{m_{e}T}{E_{\nu}^2}-(\gv +\ga )-(\gv -\ga )(1-
\frac{T}{E_{\nu}})^2\right)\nonumber  \\
~~
\eea
There are upper and lower experimental bounds for the mean square radius of 
the neutrino \cite{RPP} (90\% CL):
\be
-7.06\times10^{-11} <~ <r^2>~ < 1.26\times10^{-10} MeV^{-2}
\ee
We will restrict ourselves to positive values. In equations (5)-(8) we have
\be
\gv=-\frac{1}{2}+2~sin^2\theta_W\hspace{ 2mm},\hspace{ 8mm}\ga=-\frac{1}{2}
\ee
for $\nu=\nu_{\mu},\nu_{\tau}$,
\be
\gv=\frac{1}{2}+2~sin^2\theta_W\hspace{ 2mm},\hspace{ 8mm}\ga=\frac{1}{2}
\ee
for $\nu=\nu_e$ and we use $sin^2\theta_W=0.23$.

>From the inequality \cite{MPR}
\be
E_{\nu}\geq\frac{T+\sqrt{T^2+2m_{e}T}}{2}
\ee
and the maximum $^8$B neutrino energy \cite{Bahcall82}
\be
E_{\nu_M}=15MeV,
\ee
one can derive the lower and upper integration limits in eq.(4). These are
\be
E_{\nu_m}=\frac{T_m+\sqrt{T^2_m+2m_{e}T_m}}{2}\hspace{ 5mm},
\hspace{ 5mm}E_{\nu_M}=15 MeV
\ee
\be
T_m=E_{e_{th}}-m_e\hspace{ 5mm},\hspace{ 5mm}
T_M=\frac{2E^2_{\nu_M}}{2E_{\nu_M}+m_e}
\ee
where $E_{e_{th}}$ is the electron threshold energy in the Kamiokande 
detector.

It should be noted at this stage that the integrated cross section in (4) 
refers to a neutrino flux which is assumed to have been modified either due
to the magnetic moment spin flip inside the Sun or through flavour 
oscillations in the Sun or on its way to the detector. So an electron neutrino 
from the $^{8}B$ flux produced in the core of the Sun has a survival 
probability $\ph $ of reaching the Kamiokande detector, thus interacting 
weakly with the electron via the neutral or the charged current. The remaining
$(1-\ph)$ fraction of the flux will have oscillated to $\nu_\mu$ (or 
$\nu_\tau$) with a probability $\alpha$, thus interacting via the weak neutral
and electromagnetic currents only. Alternatively it will have 
flipped to $\nu_{e R}$ 
(or ${\nu}_{{\mu,\tau} R}$) with a probability $(1-\alpha)$ via the magnetic 
moment, thus interacting only through the electromagnetic current (see fig.1). 
The weak part of the total cross section in Kamiokande $\sigma^K_W$ may 
therefore be decomposed as follows
\bea
\sigma^K_W&=&\ph \sigma_{_W}+\alpha (1-\ph)
\sigma_{NC}\nonumber \\
&\simeq& \sigma_{_W}(0.15\alpha+\ph(1-0.15\alpha))
\eea
where $\sigma_{NC}$ denotes the weak neutral cross section and $\sigma_{_W}$ 
denotes the total $\nu_{e}e$ cross section which includes the neutral and
charged current contributions.
In eq. (16) we have used the well known fact that \cite{Okun}
\be
\sigma_{_W}\simeq6.7\sigma_{_{NC}}.
\ee
This yields the parameter $X_W$ in equation (4):
\be
X_W=0.15\alpha+\ph(1-0.15\alpha).
\ee
In order to determine $X_{int}$, we decompose the interference cross section
[eq.(8)] into its $\nu_e$ and $\nu_{\mu,\tau}$ parts, recalling as above that $\nu_e$ has 
partly survived with probability $\ph $ and partly oscillated to $\nu_{\mu}$
with probability $\alpha(1-\ph)$:
\bea
\sigma_{int}^K &= &\ph \sigma_{{\nu_e},int}+\alpha(1-\ph)
\sigma_{{\nu_{\mu},int}} \nonumber \\
&\simeq &\sigma_{{\nu_e},int}(\ph-0.37~\alpha~(1-\ph)).
\eea
In the last step we used (8), (10), (11) to obtain
\be
\frac{\sigma_{\nu_{\mu,int}}}{\sigma_{\nu_{e,int}}}\simeq-0.37
\ee
for the integrated cross sections, which yields \footnote{Since we are
interested in the upper bound for the magnetic moment which is obtained as
will be seen for vanishing charge radius, we assume 
$<r^2>_{\nu_e}=<r^2>_{\nu_{\mu,\tau}}$ and $\mu_{\nu_e}=\mu_{\nu_{\mu,\tau}}$}
\be
X_{int}=(\ph-0.37~\alpha~(1-\ph)).
\ee

If neutrinos are standard, they do not oscillate nor have any electromagnetic
properties and only the $\sigma_W$ term survives in equation (4). This 
corresponds to $X_W=1$
($\alpha=0,\ph=1$). In such a case the prediction of eq. (4) for the 
Kamiokande event rate is wrong by a solar model dependent factor $R_K$ which 
is the ratio between the data and the model prediction:
\be
S_K=R_K\int dE_{\nu}\int f(E_{\nu})\frac{d^2\sigma_W}{dT dE_{\nu}}dT.
\ee
The basic point of the paper is to equate the right hand sides of (4) and 
(22). We note that in doing so we are not merely attempting to explain the
neutrino deficit in Kamiokande which is model dependent. Even if $R_K=1$
(no neutrino deficit appears in Kamiokande) there may still be electromagnetic 
properties related to the main problem of the disappearence of the 
intermediate energy neutrinos.

Equating (4) and (22) and taking $R_K$ as an input, leaves us four parameters 
($\alpha,\ph$ and the electromagnetic ones -- $ f_{\nu}$, $<r^2>$) of which $\ph$ is
directly related to $P_{_I}$ as will be seen. We obtain
\bea
 f_{\nu}^2\!\!&=&\!\!\!\! \left( R_K-0.15~\alpha-\ph(1-0.15~\alpha~) \right)
                   \frac{\sigma_W}{B_{-EM}} \nonumber\\
     &-&\!\!<\!r^2\!>\! \left( \ph-(1-\ph)~0.37~ \alpha~ \right)
          \frac{A_{int}}{B_{-EM}}-
       \!<\!r^2\!>^2\!\frac{B_{+EM}}{B_{-EM}}
\eea
where 
\bea
\sigma_W&=&\int dE_{\nu} \int f(E_{\nu})\frac{d^2\sigma_W}{dT dE_{\nu}}dT \\
<r^2>^2B_{+EM}&=&\int dE_{\nu} \int f(E_{\nu})
\frac{d^2\sigma_{+EM}}{dT dE_{\nu}}dT \\
 f_{\nu}^2 B_{-EM}&=&\int dE_{\nu}\int f(E_{\nu})\frac{d^2\sigma_{-EM}}{dT dE_{\nu}}
dT\\
<r^2>A_{int}&=&\int dE_{\nu}\int f(E_{\nu})\frac{d^2\sigma_{int}}{dT dE_{\nu}}
dT.\\
~~\nonumber
\eea

For a given $R_K$, maximizing the magnetic moment for fixed $P_H$ ammounts to
minimizing $\alpha$ and $<r^2>$ ($\alpha =0,<r^2>=0$). This is to be expected
since it corresponds to the absence of oscillations and vanishing mean square
radius, so the neutrino deficit would only rely on the magnetic moment. 
Furthermore, it is seen that $ f_{\nu}$ also decreases with increasing  $\ph$. A
relation between $\ph$ and $P_{_I}$ can be obtained from the Chlorine data and
the solar model predictions. It involves in each model the ratio between the 
measured Chlorine event rate and its prediction, $R_{_{Cl}}$, and the quantities
$R_{_H}^{_{Cl}}, R_{_I}^{_{Cl}}$ denoting the fractions of $^{8}B$ and 
intermediate energy neutrinos ($^{7}Be$, CNO) (see table I). These quantities
are evaluated by dividing the model predicted rate for the corresponding
neutrino component by the total predicted rate in the model. We have
\be
R_{_{Cl}}=R_{_{Cl}}^{_I}P_{_I}+R_{_{Cl}}^{_H}\ph.
\ee
In fig.2 we show $\ph$  as a function of $P_{_I}$ in the four models 
considered \cite{BP95,BP92,TCL,TCCCD}.

Recent numerical analyses \cite{HR96,C97} provide us valuable information 
on the degree of suppression of the  neutrino fluxes. It is found 
that the  survival
probability of intermediate neutrinos (in the sense that their flux 
should be positive, $\phi_{Be,CNO}>0$) is in the range $2\%-4\%$. The authors 
of \cite{HR96,C97} base their analyses on the central values of the $^{8}B$
flux quoted by Kamiokande ($\phi_{_B}=(2.95\pm^{0.22}_{0.21}\pm 0.36)
\times 10^6 cm^{-2}s^{-1}$ \cite{HR96} and $\phi_{_B}=(2.73\pm 0.17\pm 0.34)
\times 10^6 cm^{-2}s^{-1}$ \cite{C97}), while the recent values quoted by
SuperKamiokande are manifestely lower. They obtained the fits
\bea
\rm {ref.17)}~~~~~{\phi}_{Be+CNO}&=&(-2.5\pm 1.1)\times 10^9 
cm^{-2} s^{-1}\\
\rm {ref.18)}~~~~~{\phi}_{Be+CNO}&\leq&0.7\times 10^9 cm^{-2} s^{-1}~~~ 
(3\sigma)
\eea
which, compared with the theoretical predictions for eight solar models
\cite{BP95} - \cite{C97} (see table II), gives
\be
P_{_I} (3\sigma , \rm {all~eight~models})\leq 0.18.\\
\ee

A straightforward argument which is practically independent of any solar 
model assumptions shows that by decreasing $\phi_{_B}$ $(=\phi^{Kam}_{_B})$,
the flux of the intermediate energy neutrinos is further decreased. In fact, 
using the equations \cite{C97}
\bea
S_{_{Ga}}&=&\sum_{i} \sigma_{_{Ga,i}}\phi_{_i}~~~(i=pp,pep,^7\!\!{Be},CNO,
^8\!\!{B})\\
S_{_{Cl}}&=&\sum_{j} \sigma_{_{Cl,j}}\phi_{_j}~~~(j=^7\!\!{Be},CNO,^8\!\!{B})
\eea
together with the luminosity constraint \cite{BP95,C97} ($L_{\odot}=
1.367\times 10^{-1}Wcm^{-2}$)
\be
L_{\odot}=\sum_{k}(\frac{Q}{2}-\!<\!E_{\nu}\!>_k\!)\phi_k~~~(k=pp,pep,
^7\!\!{Be},CNO,^8\!\!{B})
\ee
with $Q=26.73 MeV$ (total energy released in each neutrino pair production)
and taking $\phi_{pep}=0.0021 \phi_{pp}$, one gets upon elimination of $\phi_{pp}$:
\bea
\phi_{Be}&=&10.4\phi_B-28.8~~,\\
\phi_{CNO}&=&\!\!-8.46\phi_B+22.2 ~~.
\eea
In these equations $\phi_B$ is given in units of $10^6 cm^{-2}s^{-1}$ while 
$\phi_{Be}$, $\phi_{CNO}$ are given in units of $10^9 cm^{-2}s^{-1}$ and
we used $ S_{_{Cl}}=2.55\pm 0.25SNU$ \cite{Lande96} and the 
weighted average of SAGE \cite{Abdurashitov} and Gallex \cite{Anselmann} data,
$\bar {S}_{_{Ga}}=73.8\pm 7.7SNU$. Since $\rm \phi_{Be}$ is of the order
of 5 times $\rm \phi_{CNO}$ or larger, it is obvious from (35), (36) that a
decrease in $\rm \phi_{B}^{Kam}$ leads to a decrease in the total flux
$\rm \phi_{Be+CNO}$. We therefore conclude that SuperKamiokande data 
strengthen the general belief that the intermediate energy neutrinos (mainly
$^7\!\!{Be}$) are strongly suppressed. Thus it appears from eqs. (29), (30) 
that to within 97\% CL at least, one can take a vanishing $P_{_I}$.

Clearly if the neutrino is endowed with electromagnetic properties, it may in
principle have a magnetic moment $\mu_{\nu}$ and a mean square radius 
$<\!r^2\!>$. The upper bound on the magnetic moment is obtained from eq. (23)
in the limiting situation of vanishing $<\!r^2\!>$ and $\alpha$ (absence of 
flavour oscillations). We are bound to restrict ourselves to solar models 
\cite{BP95,BP92,TCL,TCCCD} for which the ratios 
$R_{_I}^{_{Cl}}$ and $R_{_H}^{_{Cl}}$ are
available and with comparatively small error bars. Using eqs. (23) and (28)
we display the neutrino magnetic moment in Bohr magnetons $ f_{\nu}$ against 
$P_{_I}$, the survival probability for intermediate energy neutrinos, up to
$P_{_I}=0.18$ (see eq. (31)) and with $<\!r^2\!>=0, \alpha=0$. We consider two
situations: in fig. 3 we use the preliminary results \cite{SuperKamiokande}
from SuperKamiokande, $\phi^{_{SK}}_{_B}=(2.51\pm^{0.14}_{0.13}\pm 0.18)
\times 10^6 cm^{-2}s^{-1}$ with a threshold $ E_{e_{th}}=7.0MeV$ and in fig. 4
we use the Kamiokande results \cite{Kamiokande}, 
$\phi^{_{Kam}}_{_B}=(2.80\pm 0.19
\pm 0.33)\times 10^6 cm^{-2}s^{-1}$ and threshold $E_{e_{th}}=7.0MeV$. We
also show in figs. 5, 6 the behaviour of $\mu_{\nu_{e}}$ as a function of 
$<r^2>$ in the limit $\alpha=0$ and as a function of $\alpha$ in the limit 
$<r^2>=0$ respectively.

It is clear that the results for the upper bounds on $\mu_{\nu_e}$ obtained 
using SuperKamiokande are not only stricter than the ones using Kamiokande, 
but their spread for the different models is also much smaller. As shown 
above, up to more than $2\sigma$ one can take $P_{_I}=0$, so it is appropriate 
to consider the left ends of these curves as the actual upper limits on 
$\mu_{\nu_e}$ from experiment and theoretical models. We have in these 
conditions
\bea
\mu_{\nu_e}&\leq&(2.18-2.29)\times 10^{-10}\mu_{_B}~~\rm {SuperKamiokande}\\
\mu_{\nu_e}&\leq&(3.46-4.13)\times 10^{-10}\mu_{_B}~~\rm {Kamiokande}.
\eea
These may increase by approximately 50\% if one relaxes the constraint of a 
vanishing $P_{_I}$ and let it approach its $3\sigma$ upper limit of 0.18
(see figs. 3, 4). We
also note that the disparities on the predictions for the $^8{B}$ flux among 
solar models (table I), related to uncertainties in the astrophysical factor
$S_{17}$, are hardly reflected on the upper bound on $\mu_{\nu}$ for all neutrino types.

An essential development which may further improve the bound (37) is the 
decrease in $E_{e_{th}}$, the recoil electron threshold energy in 
$\nu_{_{e,x}}e$ scattering. This decrease implies a
decrease in the ratio of integrals $\sigma_{_W}/B{_{-EM}}$ appearing in
equation (23). This is related to the fact that for decreasing energy and a
sizeable neutrino magnetic moment, the electromagnetic contribution to the
scattering increases faster than the weak one. The above referred ratio of
integrals leads through (23) for constant values of $R_{_K}$ and $P_{_H}$
to a decrease in the upper bound for $f_{\nu}$. Both the Kamiokande and
the SuperKamiokande detector so far operate with a threshold of 7.0 MeV.
The SuperKamiokande collaboration plans to improve their threshold down to 
5.0 MeV in the near future. The forthcoming SNO experiment \cite{SNO} also
aims to operate near this threshold. For $E_{e_{th}}$ =5.0 MeV
and the same ratio of data/model prediction for the
$^8{B}$ neutrino flux ($R_{_K}$), the bound (37) would be 
decreased to $(1.6-1.7)\times 10^{-10} \mu_{_B}$. Hence a further decrease 
in the electron threshold energy will be a welcome improvement.

\vspace {10 mm}

{\bf \large 3. Conclusions}

\vspace {6 mm}

We have investigated the existence of an upper bound on the electric neutrino
magnetic moment $\mu_{\nu_e}$ from solar neutrino experiments. Besides 
laboratory bounds, this looks a promising source for constraining 
all neutrino magnetic moments and thus establishing upper limits 
on these quantities. The 
strictest laboratory bounds existent up to date refer to electron 
anti-neutrinos ($\mu_{\bar \nu_e}<1.8\times 10^{-10}\mu_{_B}$ 
\cite{Derbin}) and a new experiment \cite{NIM} 
aimed at providing new constraints is expected to start operation soon. 
Regarding laboratory bounds on $\mu_{\nu_e}$, the limit is higher: 
$\mu_{\nu_e}<10.8\times 10^{-10}\mu_{_B}$ \cite{RPP}. We believe the 
present work, where we used SuperKamiokande data, provides a new bound 
on  $\mu_{\nu_e}$ of the order of the one available on $\mu_{\bar \nu_e}$. 
We find $\mu_{\nu_e}<(2.2-2.3)\times 10^{-10}\mu_{_B}$. This was derived on
the assumption of equal neutrino magnetic moments for different flavours
and  a total suppression of intermediate energy neutrinos: 
$P_{_I}=0$.

>From the theoretical standpoint, the uncertainties in $S_{17}$, the parameter
describing the $^8{B}$ flux prediction, although not irrelevant, do not play
a crucial role. In fact, the upper bound on $\mu_{\nu_e}$ is only very 
moderately sensitive to them.

On the other hand, the decrease in the recoil electron threshold energy in the
solar neutrino electron scattering may further constrain this bound. Thus not
only the expected improvement in SuperKamiokande, but also the SNO experiment
\cite{SNO} examining this process with a 5 MeV threshold or possibly lower 
will be essential for the purpose.

\newpage

\begin{center}
\begin{tabular}{|c|c|c|c|c|c|c|} \hline
Model&$R_{_{Cl}}^{_I}$&$R_{_{Cl}}^{_H}$&$R_{_{Cl}}$&$\phi_{_B}$&$R_{_{SK}}$&
$R_{_K}$ \\ \hline
BP95 \cite{BP95}&0.209&0.791&0.274&6.62&0.379&0.423 \\ \hline
BP92 \cite{BP92}&0.221&0.777&0.319&5.69&0.441&0.492 \\ \hline
TCL \cite{TCL}&0.248&0.752&0.401&4.43&0.567&0.632 \\ \hline
TCCCD \cite{TCCCD}&0.292&0.706&0.443&3.8&0.661&0.737 \\ \hline
\end{tabular}

\vspace{ 10mm}
\end{center}
Table I - The columns $R_{_{Cl}}^{_I}, R_{_{Cl}}^{_H}, R_{_{Cl}},\phi_{_B},
R_{_{SK}},R_{_K}$ denote respectively the fractions of intermediate and high 
energy neutrinos in the Chlorine experiment, the ratio of the total measured
signal and the model prediction, the $^{8}B$ flux prediction and the ratio
data/model prediction for the SuperKamiokande and Kamiokande data in each 
of the four models \cite{BP95} - \cite{TCCCD}. Units of $\phi_{_B}$ are in
$10^6 cm^{-2}s^{-1}$.

\vspace{ 15mm}

\begin{center}
\begin{tabular}{|c|c|} \hline
Model&$\phi_{_{Be+CNO}}(\times 10^9cm^{_{-2}}s^{_{-1}})$ \\ \hline
BP95 \cite{BP95}&6.31 \\ \hline
BP 92 \cite{BP92}&5.81 \\ \hline
TCL \cite{TCL}&5.37 \\ \hline
TCCCD \cite{TCCCD}&4.94 \\ \hline
P94 \cite{PRO94}&6.38 \\ \hline
DS96 \cite{DS96}&4.47 \\ \hline
RVCD96 \cite{RVCD96}&5.84 \\ \hline
FRANEC96 \cite{CIR97}&5.47 \\ \hline
\end{tabular}
\end{center}
\vspace{ 10mm}
Table II - The flux of $^7{Be}$ and CNO neutrinos in each of the eight models
considered in eq. (31).

~\newpage

\begin{figure}
\begin{picture}(18,20)
\put(1,2){\epsfig{figure=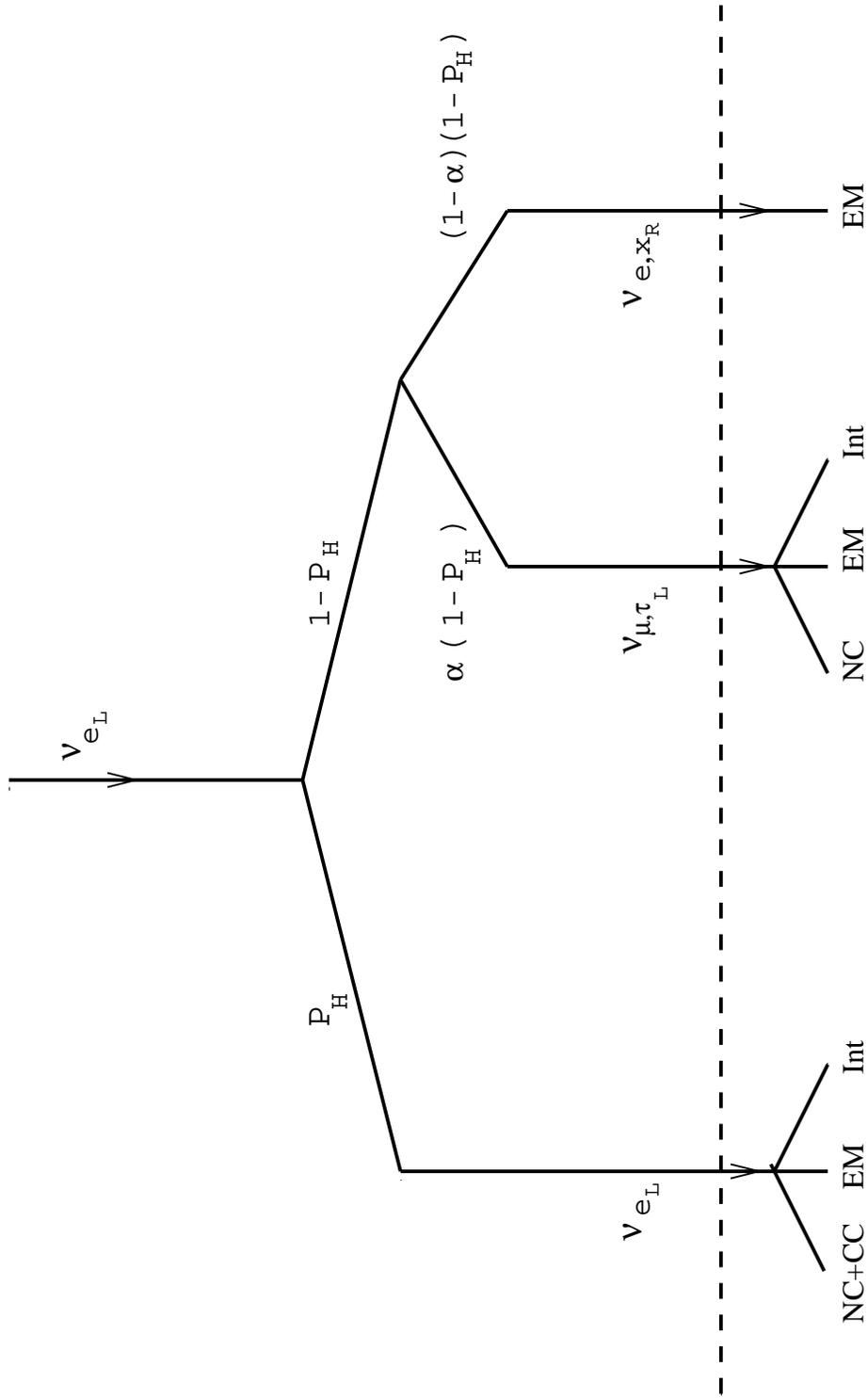,width=15cm}}
\end{picture}
\caption{A fraction $P_H$ of the initial $\nu_{eL}$ flux remains unaltered and
interacts with $e^-$ in Kamiokande. Its cross section contains a weak 
contribution (charged (CC) and neutral current (NC)), an electromagnetic one 
and the interference between them. Of the remaining $(1-P_H)$, a fraction 
$\alpha$ is converted to $\nu_{\mu,{\tau}L}$ and interacts without the weak 
charged current while the remaining $(1-\alpha)(1-P_H)$ interacts only 
electromagnetically.}
\end{figure}

\begin{figure}
\begin{picture}(18,20)
\put(1,2){\epsfig{figure=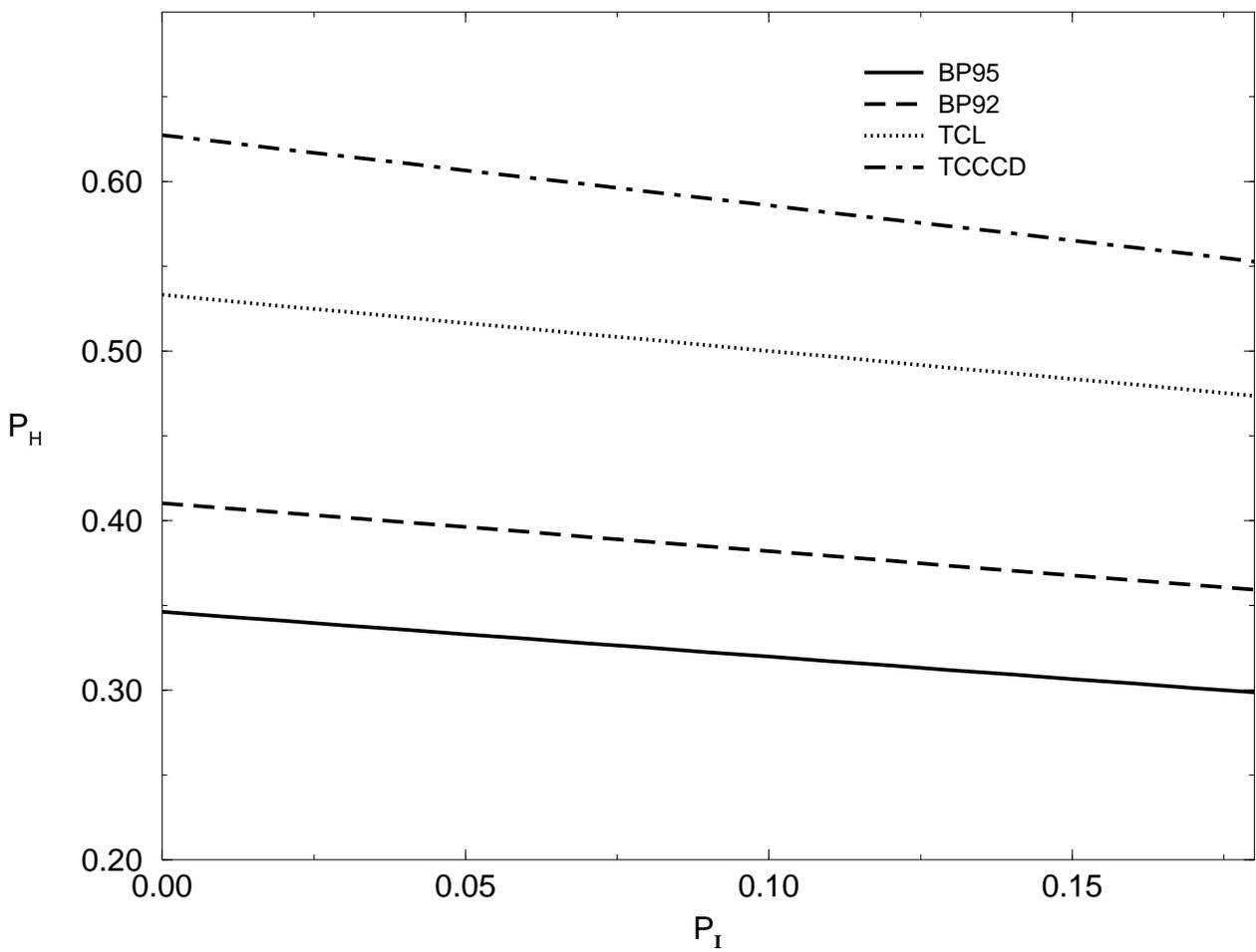,width=15cm}}
\end{picture}
\caption{$P_H$ as a function of $P_I$ up to $P_I=0.18$ for each of the four 
models \cite{BP95}-\cite{TCCCD}.}
\end{figure}

\begin{figure}
\begin{picture}(18,20)
\put(1,2){\epsfig{figure=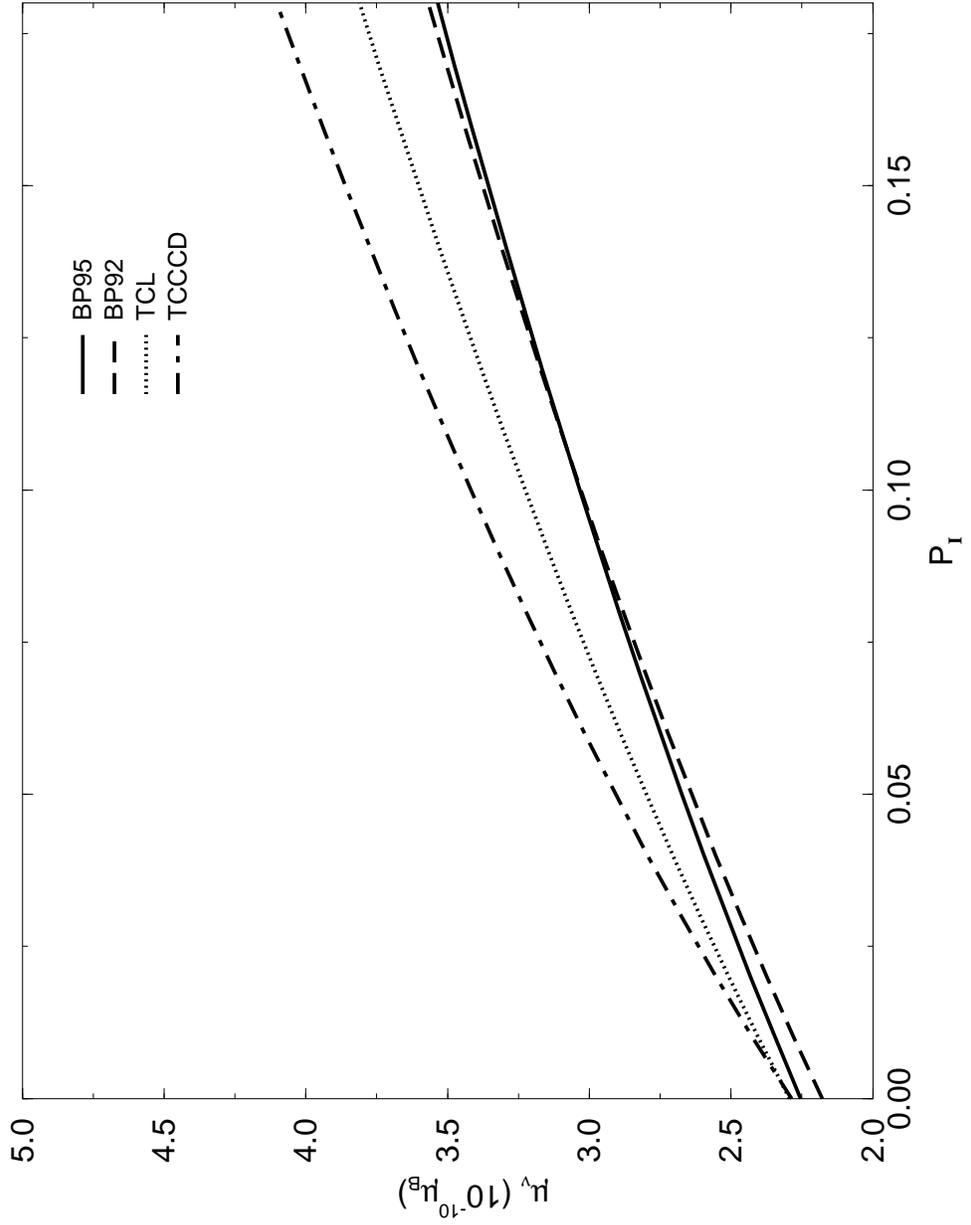,width=15cm}}
\end{picture}
\caption{The neutrino magnetic moment as a function of $P_I$ for ${\alpha}=0$,
$<r^2>=0$ in each of the four models \cite{BP95}-\cite{TCCCD}. Within 97\% CL
at least, one can take $P_I=0$, so the corresponding upper bound on $\mu_{\nu_e
}$ is in each model the value at the left end of the curve. The experimental 
data used are from SuperKamiokande.}
\end{figure}

\begin{figure}
\begin{picture}(18,20)
\put(1,1){\epsfig{figure=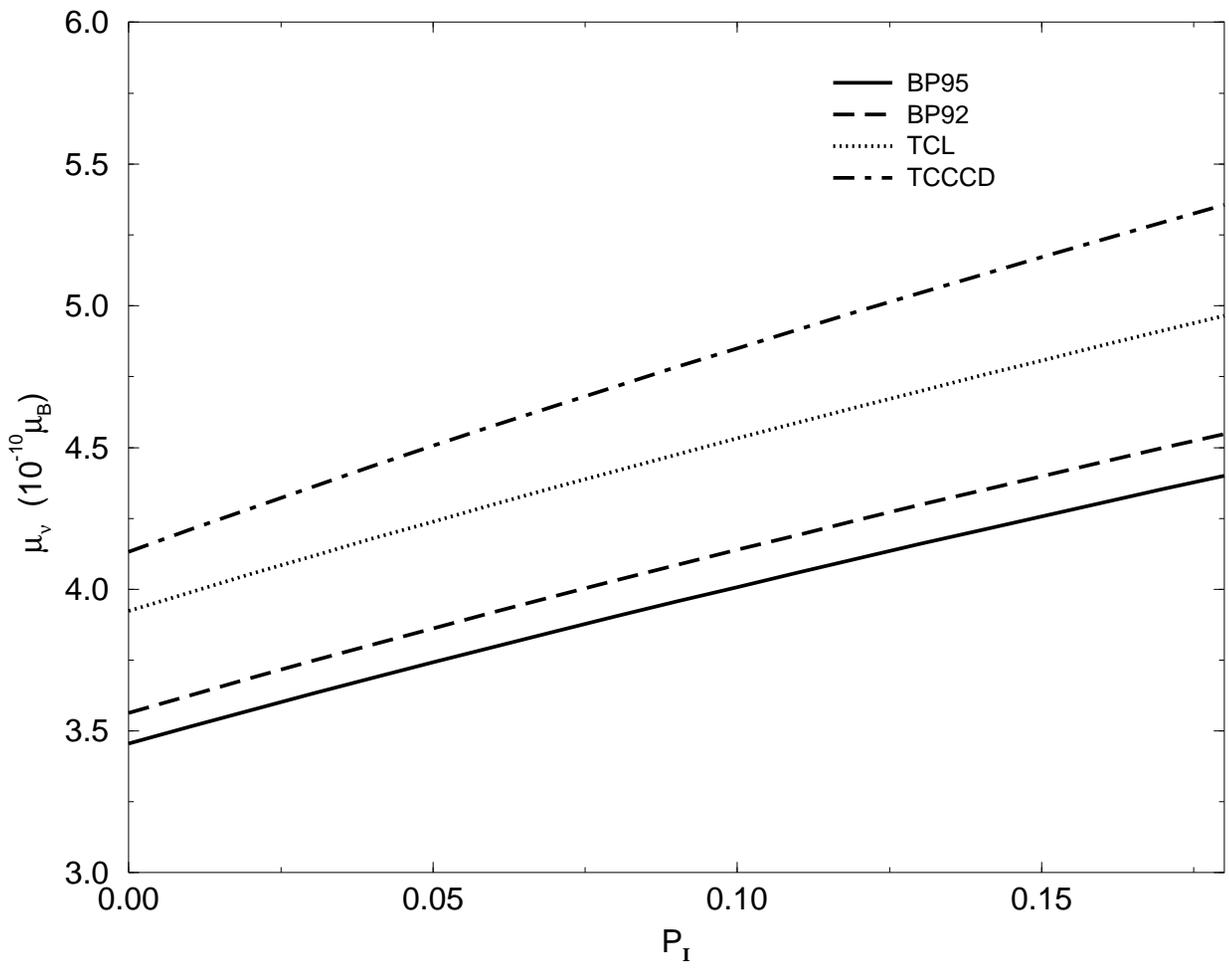,width=15cm}}
\end{picture}
\caption{Same as fig. 3 with Kamiokande data.}
\end{figure}

\begin{figure}
\begin{picture}(18,20)
\put(1,2){\epsfig{figure=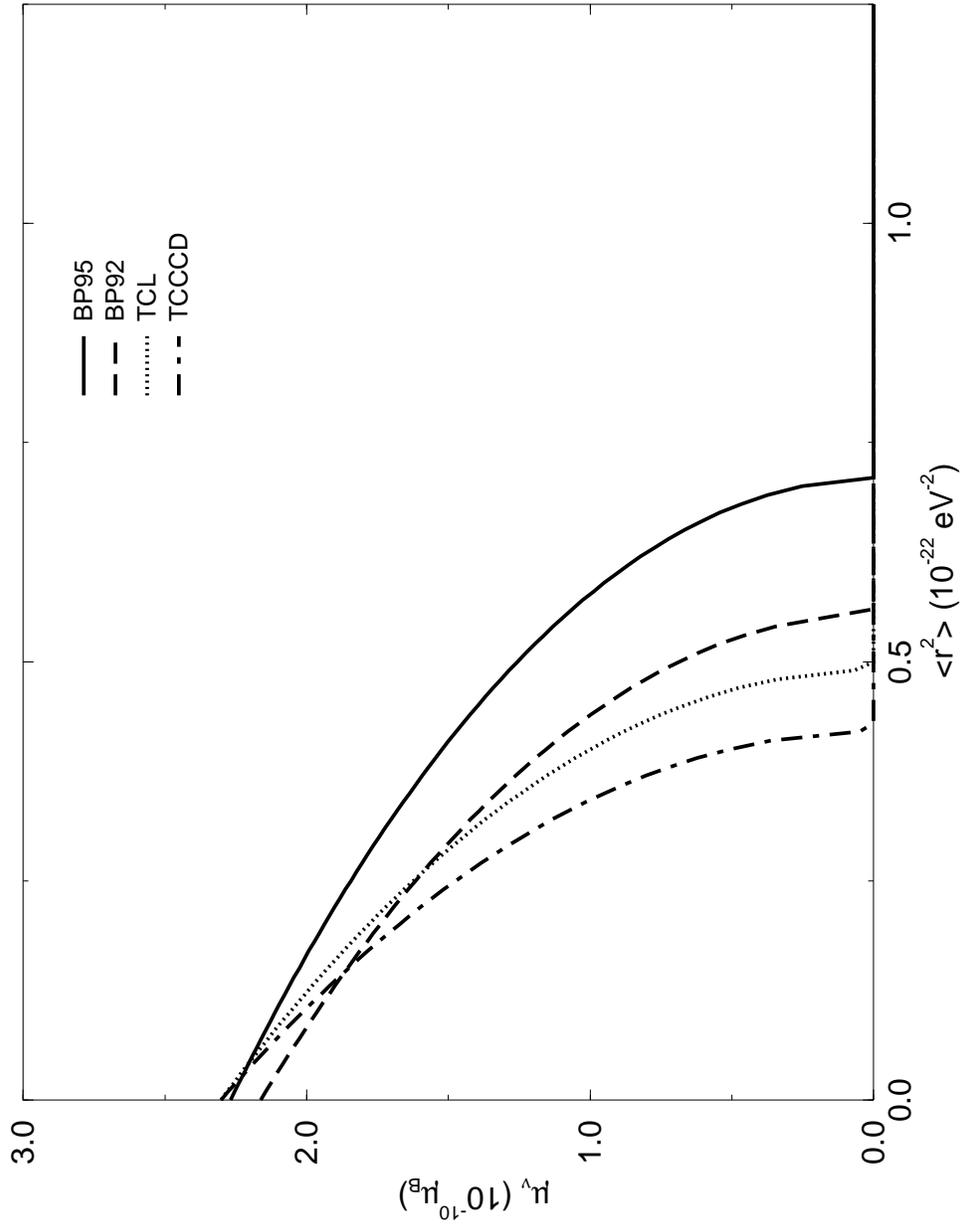,width=15cm}}
\end{picture}
\caption{Neutrino magnetic moment as a function of the mean square radius
$<r^2>=0$ ion the limit $\alpha=0$ and in each of the four models 
\cite{BP95}-\cite{TCCCD}.
The upper bound on $\mu_{\nu_e}$ is in each model the left end of
the curve.}
\end{figure}

\begin{figure}
\begin{picture}(18,20)
\put(1,1){\epsfig{figure=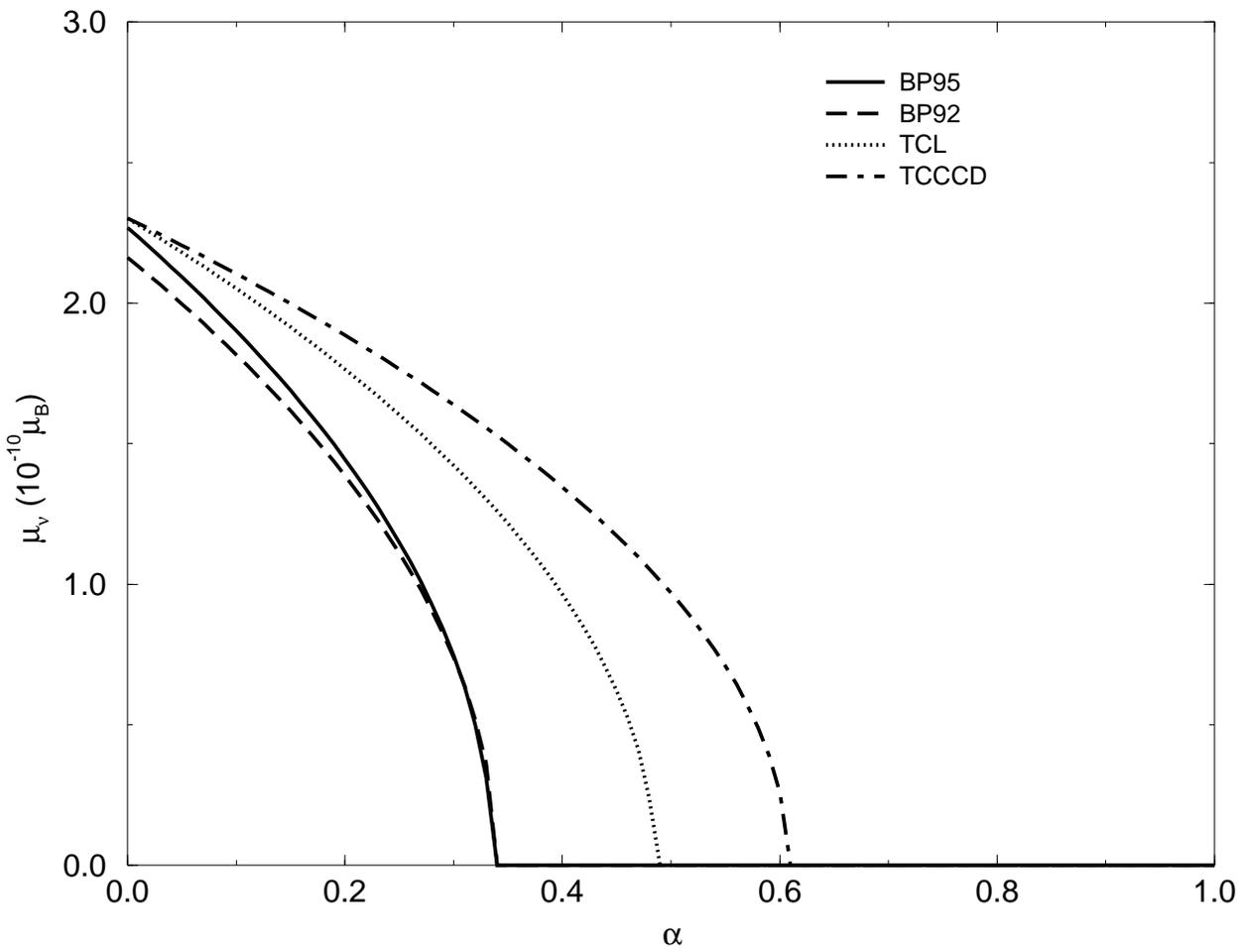,width=15cm}}
\end{picture}
\caption{Same as fig. 5 as a function of $\alpha$ in the limit $<r^2>=0$.}
\end{figure}

\end{document}